\documentclass[a4paper]{jpconf}
\usepackage{graphicx,cite,amsmath}
\begin{document}
\title{Stability of particles in two-dimensional quasicrystals against phasonic perturbations}

\author{M Martinsons$^1$ and M Schmiedeberg$^1$}

\address{$^1$Institut f\"ur Theoretische Physik 1, Friedrich-Alexander-Universit\"at Erlangen-N\"urnberg, Staudtstr. 7, D-91058 Erlangen, Germany}

\ead{michael.schmiedeberg@fau.de}

\begin{abstract}
  We investigate particles in two-dimensional quasicrystalline interference patterns and present a method to determine for each particle at which phasonic displacement a phasonic flip occurs. By mapping all particles into characteristic areas of reduced phononic and phasonic displacements, we identify the particles that are close to edges of these areas and can easily flip. In contrast, the particles in the center are hardly affected by phasonic fluctuations. Our results are important e.g. for light-induced colloidal structures or cold atomic gases in laser traps. In addition, our approach can help to predict how thermal fluctuations induce phasonic flips in intrinsic quasicrystals with structures close to interference patterns.
\end{abstract}

\section{Introduction}

Quasicrystals -- well-ordered structures that lack periodic translational symmetry \cite{shechtman84,levine84} -- possess additional degrees of freedom termed phasons. In the long-wavelength limit phasons do not cost any free energy \cite{bak85,bak85b,socolar86} and correspond to correlated rearrangements of particles \cite{socolar86,kromer12,sandbrink13,kromer13,martinsons14}. At a short wavelength they are excitations known as phasonic flips \cite{socolar86}. In this article we determine for all particles in typical quasicrystals at which phasonic displacement a flip occurs.

In order to demonstrate our approach, we employ quasicrystalline patterns that are connected to the patterns of interfering laser beams with quasicrystalline symmetry \cite{kromer12,sandbrink13,kromer13,martinsons14,burns90,guidoni97,guidoni99,sanchez05,ablowitz06,gorkhali06,freedman06,freedman07,schmiedeberg07,schmiedeberg08,mikhael08,mikhael10,schmiedeberg10,schmiedeberg12,neuhaus13a,jagannathan13,ruehle15,viebahn19}. Laser interference patterns have been used to induce desired structures in colloidal suspensions \cite{burns90,schmiedeberg08,mikhael08,mikhael10,schmiedeberg10,ruehle15}, liquid crystals \cite{gorkhali06}, photosensitive materials \cite{freedman06,freedman07}, or cold atom systems \cite{guidoni97,guidoni99,sanchez05,ablowitz06,jagannathan13,viebahn19}. Such systems can be employed to study the competition of incompatible symmetries \cite{mikhael08,schmiedeberg10,schmiedeberg06,neuhaus13,sagi16}, the differences between various rotational symmetries \cite{mikhael10,schmiedeberg12}, the dynamics of particles \cite{guidoni99,schmiedeberg07}, the propagation of defects \cite{freedman07,sandbrink14}, and the growth process on incommensurate substrates \cite{neuhaus13a,neuhaus14,sandbrink14}. In all of these cases the maxima of the intensity pattern can be very diverse both in intensity and in shape. As a consequence, some particles are strongly trapped while other particles can easily jump to another position \cite{schmiedeberg07}, such that thermally-induced phasonic flipping can be observed for only some particles \cite{schmiedeberg08,mikhael08,schmiedeberg10}.

\section{Method and background: laser patterns, phasons and characteristic areas}
\label{sec:methods}

The interference of $N$ symmetrically-arranged laser beams polarized in the same direction leads to an intensity field \cite{gorkhali06,schmiedeberg07,schmiedeberg12} $I\left(\vec{r}\right)\propto  \sum\nolimits_{j=0}^{N-1} \sum\nolimits_{k=0}^{N-1} \textrm{cos}\left[(\vec{G}_{j} - \vec{G}_{k})\cdot \vec{r} + \phi_{j} - \phi_{k}\right]$, where $\vec{G}_{j}=2\pi/a_V(\cos[2\pi j/N],\sin[2\pi j/N])$ with a length $a_V$ are the projections of the laser wave vectors on the sample plane and $\phi_{j}$ are the phases of the beams as specified in the next paragraph. Examples for various $N$ are shown in \cite{schmiedeberg12}. The forces acting on a particle in such a laser field are approximately given by the average of the intensity over the trapped particle \cite{hanes12,hanes13} resulting in various possible effective potential energy landscapes \cite{ruehle15}. Here we consider point particles.

\begin{figure}
\centering
\includegraphics[width=\textwidth]{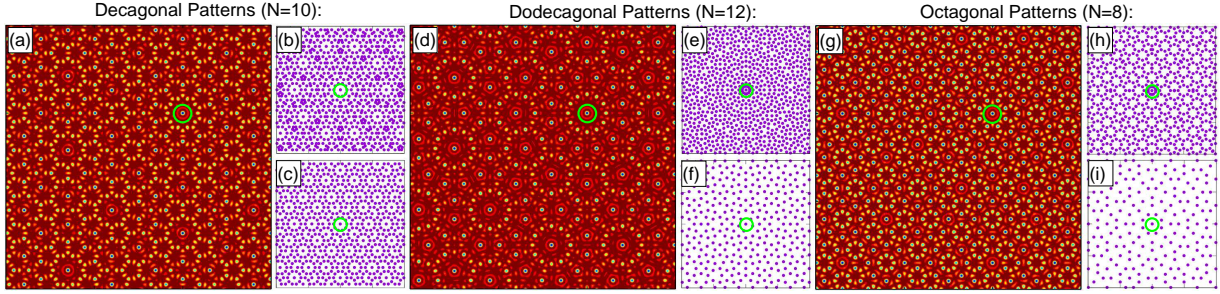} 
\caption{(a,d,g) Intensity patterns obtained by $N=10$, $12$ or $8$ interfering laser beams. The patterns are decorated by particles where (b,e,h) all local maxima are occupied or (c,f,i) only the most pronounced maxima leading approximately to (c) a Tuebingen tiling, (f) a shield tiling or (i) an Ammann-Beenker tiling. Green circles with radius $a_V$ mark the global symmetry centers.}
\label{fig:fig1}
\end{figure}

Interference patterns can be decorated in various ways, e.g. by occupying all local intensity maxima or only the most pronounced ones such that approximately typical tilings are obtained (see fig.~\ref{fig:fig1}). Such tilings are found due to the similarity of selecting pronounced maxima in laser fields and selecting particles in superspace by an acceptance window \cite{jagannathan13,sandbrink14,hielscher17}.

For quasicrystals of rank $D=4$ as considered here the phases $\phi_{j}$ of the laser beams can be decomposed into phononic displacements $\vec{u}=(u_x,u_y)$ and phasonic displacements $\vec{w}=(w_{x},w_{y})$: $\phi_{j} = \vec{G}_{j} \cdot \vec{u} + \vec{G}_{kj\, \textrm{mod}\, N} \cdot \vec{w}$ \cite{socolar86,sandbrink13}, where, e.g.  $k=3$ for $N=5$ or $N=8$ and $k=5$ for $N=12$ \cite{socolar86,sandbrink13}. Note, for quasicrystals with $D>4$ there are additional phasonic variables \cite{martinsons14}.

Some combinations of phononic and phasonic displacements $\Delta \vec{u}$ and $\Delta \vec{w}$ do not change the intensity field \cite{kromer12,sandbrink13,martinsons14}. Every particle within the quasicrystal can be mapped inside so-called characteristic areas by repeatedly displacing it by $\Delta \vec{u}$ and $\Delta \vec{w}$ until the particle is as close to the origin in the $u_x$-$u_y$-$w_x$-$w_y$-space as possible \cite{kromer12,sandbrink13,martinsons14}. More details on $\Delta \vec{u}$, $\Delta \vec{w}$ and the characteristic areas are given in \cite{kromer12} for decagonal and in \cite{sandbrink13} for dodecagonal or octagonal systems.

\section{Results}\label{sec:results}

\begin{figure}
\centering
\includegraphics[width=\textwidth]{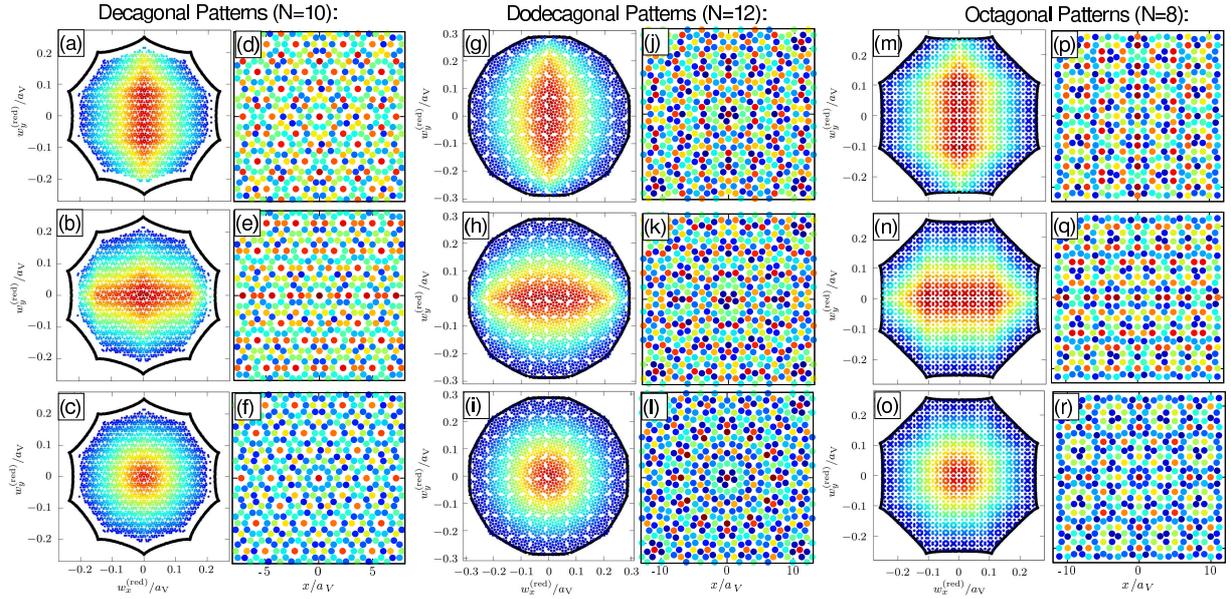} 
\caption{Particles in the $w_x$-$w_y$-plane mapped into the characteristic areas (black lines). We employ (a,b,c) $2728$ particles in a Tuebingen tiling, (g,h,i) $2170$ in a shield tiling and (m,n,o) $2204$ in an Ammann-Beenker tiling. Colors indicate the minimal amplitude that is needed to obtain a phasonic flip for fluctuations in (a,g,m) $x$-direction, (b,h,n) $y$-direction and (c,i,o) an arbitrary direction. (d,e,f,j,k,l,p,q,r) Original positions of the particles with the same colors.}
\label{fig:fig2}
\end{figure}

First, all particles are mapped into the characteristic areas. Examples of such areas in the $w_x$-$w_y$-plane are shown in figs.~\ref{fig:fig2}(a,b,c,g,h,i,m,n,o) (surrounded by black lines). Then, one can determine for each particle what phasonic displacement is necessary to obtain a phasonic flip. This phasonic displacement is given by the distance to the boundary of the characteristic area, because a particle that in the $w_x$-$w_y$-plane is moved out of that area has to be displaced by $\Delta \vec{u}$ and $\Delta \vec{w}$ in order to be mapped back into the characteristic area \cite{kromer12,sandbrink13,martinsons14}. $\Delta \vec{u}$ corresponds to the jump associated with a phasonic flip of the particle \cite{kromer12,sandbrink13,martinsons14}.

For the decagonal Tuebingen tiling, the dodecagonal shield tiling and the octagonal Ammann-Beenker tiling, we determined the distance of each particle from the boundary of the characteristic area in the cases of phasonic displacements only in $x$-direction (first line in fig.~\ref{fig:fig2}), only in $y$-direction (second line in fig.~\ref{fig:fig2}) or in an arbitrary direction (bottom line in fig.~\ref{fig:fig2}). The particles are also shown in their original arrangement in figs.~\ref{fig:fig2}(d,e,f,j,k,l,p,q,r) where the colors are the same as in figs.~\ref{fig:fig2}(a,b,c,g,h,i,m,n,o) and indicate the respective phasonic distances until flipping. Therefore, blue particles already flip if there is just a small phasonic displacement, e.g. due to small thermal fluctuations. Particles in red are unlikely to flip, i.e. they are the most stable particles in the tilings.

\begin{figure}
\centering
\includegraphics[width=\textwidth]{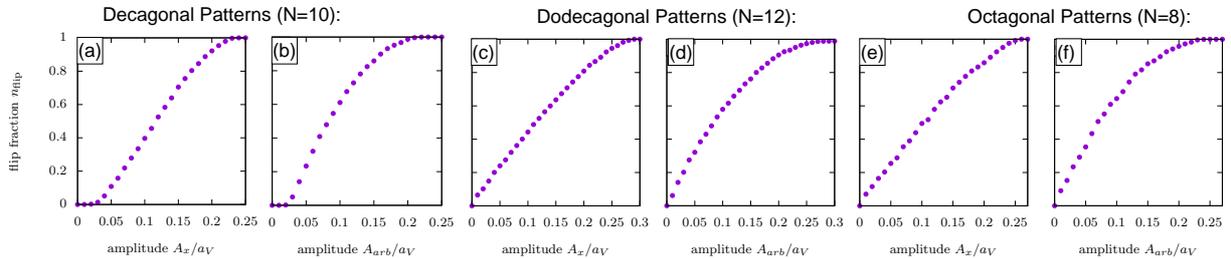} 
\caption{Fraction $n_{\mathrm{flip}}$ of flipping particles as a function of the amplitude if the phasonic fluctuations occur along (a,c,e) the $x$-direction or (b,d,f) an arbitrary direction.}
\label{fig:fig3}
\end{figure}

In fig.~\ref{fig:fig3} we show the fraction of particles that are flipping for a given phasonic displacement. Interestingly, we observe an almost linear behavior in case of small phasonic displacements in dodecagonal or octagonal tilings. However, in the decagonal Tuebingen tiling a certain phasonic displacement is needed in order to obtain a phasonic flip at all. The reason is that particles in weak maxima of the laser field were left out when constructing the Tuebingen tiling. 

\section{Conclusions}
\label{sec:conclusions}

We described a method how the stability of particles in quasicrystalline structures related to interference patterns can be predicted, i.e. we can determine which particles flip for a given phasonic displacement that might e.g. be caused by thermal fluctuations. We presented the results of our approach for three examples, namely the decagonal Tuebingen tiling, the dodecagonal shield tiling and the octagonal Ammann-Beenker tiling. We found that the particles in a Tuebingen tiling are all stable with respect to small phasonic fluctuations while in the other cases flips occur already for very small phasonic displacements. Our method can also be applied to other quasicrystalline structures. This obviously includes all structures induced by laser fields \cite{kromer12,sandbrink13,kromer13,martinsons14,burns90,guidoni97,guidoni99,sanchez05,ablowitz06,gorkhali06,freedman06,freedman07,schmiedeberg07,schmiedeberg08,mikhael08,mikhael10,schmiedeberg10,schmiedeberg12,neuhaus13a,jagannathan13,ruehle15,viebahn19}, but also quantum quasicrystals \cite{gopalakrishnan14,sandbrink14b,lifshitz14,han18} have been shown to be similar \cite{sandbrink14b}. Furthermore, even intrinsic quasicrystals and their thermally fluctuating phasonic displacements can be analyzed by comparing them to laser fields \cite{hielscher17}. Note that phasonic flips due to thermal fluctuations are also important for the growth process of intrinsic quasicrystals \cite{sandbrink14,achim14,schmiedeberg17,martinsons18,gemeinhardt18} and in future might be used to predict which parts of a quasicrystal grow without flips and which particles are more likely placed in a wrong position during the growth process.

\ack
The work was supported by the German Research Foundation (Schm 2657/2 and Schm 2657/4).

\end{document}